\documentclass[apj]{emulateapj}

\shorttitle{Discovery of the radio and X-ray counterpart of the HESS J1731-347}
\shortauthors{Tian et al.}

\begin{document}

\title{Discovery of the radio and X-ray counterpart of TeV $\gamma$-ray source HESS J1731-347} 

\author{W.W. Tian\altaffilmark{1,2}, D.A. Leahy\altaffilmark{2}, M. Haverkorn\altaffilmark{3}, B. Jiang\altaffilmark{4,5}}
\altaffiltext{1}{National Astronomical Observatories, CAS, Beijing 100012, China; tww@bao.ac.cn} 
\altaffiltext{2}{Department of Physics \& Astronomy, University of Calgary, Calgary, AB, T2N 1N4, Canada; tww@iras.ucalgary.ca}
\altaffiltext{3}{Jansky Fellow, National Radio Astronomy Observatory and
Astronomy Department, University of California-Berkeley, 601 Campbell Hall, Berkeley CA 94704, USA}
\altaffiltext{4}{Department of Astronomy, University of Massachusetts, 710 North Pleasant Street, Amherst, MA 01003, USA}
\altaffiltext{5}{Department of Astronomy, Nanjing University, Nanjing 210093, P.R.China}
\begin{abstract}
We discover a faint shell-type radio and X-ray source, G353.6-0.7, associated with HESS J1731-347. G353.6-0.7 is likely an old supernova remnant (SNR), based on radio (0.8 GHz, 1.4 GHz and 5 GHz), infrared (8 $\mu$m from the GLIMPSE Legacy Project and 21 $\mu$m from the Midcourse Space Experiment), and X-ray (0.1 keV - 2.4 keV from the ROSAT survey and 5 - 20 keV from the INTEGRAL survey) data. The SNR, centered at ({\sl l}, {\sl b})=(353.55, -0.65) with a radius of $\sim$ 0.25$^{\circ}$, closely matches the outline of the recently discovered extended TeV source HESS J1731-347, which has no previously identified counterpart. A diffuse X-ray enhancement detected in the ROSAT all-sky survey is coincident with lower half shell of the SNR. Therefore the SNR is the best radio counterpart of both the HESS source and the diffuse X-ray enhancement. G353.6-0.7 has an age of $\sim$ 27000 yrs.  Altogether, the new discovery provides the best case that an old SNR emits TeV $\gamma$-rays.
:   
\end{abstract}


\section{Introduction}
About 52 Galactic TeV $\gamma$-ray sources have been detected by ground-based high energy telescopes (e.g. the High Energy Stereoscopic System (HESS), Aharonian et al. 2006; the MAGIC telescopes, Albert et al. 2006) so far. Some of them do not have counterparts in other wavebands. Seeking their Galactic counterparts in other wavebands could play a key role for us to understand the origin of these sources and distinguish different acceleration mechanisms to emit TeV $\gamma$-rays. Among identified counterparts, young shell-type supernova remnants (SNRs) are one of the four Galactic populations to generate very high energy (VHE) $\gamma$-rays, i.e. pulsar wind nebulae (PWN), X-ray binaries, shell SNRs and a young stellar cloud Westerlund 2 (Aharonian et al. 2007, Landi et al. 2006, Torres et al. 2003, Enomoto et al. 2002). Recently, the old SNR W41 has been proposed to associate with the extended TeV source HESS J1834-087 based on observations (Tian et al. 2007; Leahy \& Tian 2008), which is theoretically expected (Yamazaki et al. 2006, Fang \& Zhang 2008). However, this is the only likely association of an old SNR with a TeV source up to now. We report here the discovery of a new faint radio and X-ray shell which closely matches the extended TeV source HESS J1731-347 without previously identified counterpart (Aharonian et al. 2008). We use new radio, infrared and X-ray data to identify the shell as an old SNR, therefore obtaining second association of a TeV source with an old SNR.   

\section{Radio and X-ray Observations}
The 1420 MHz radio continuum and 21 cm HI emission data sets come from the Southern Galactic Plane Survey (SGPS) which is described in detail by Haverkorn et al. (2006) and McClure-Griffiths et al. (2005).  The SGPS images
the HI line emission and 1420~MHz continuum in the range 253$^{\circ} <
l < 357^{\circ}$ and $|b| < 1.5^{\circ}$, mostly in the fourth quadrant of our Galaxy,  with the
Australia Telescope Compact Array (ATCA) and the Parkes 64m single
dish telescope. The continuum observations have a resolution of
100~arcsec and a sensitivity better than 1~mJy/beam. The HI data have
an angular resolution of 2~arcmin, a rms sensitivity of $\sim$1~K and
a velocity resolution of 1~km~s$^{-1}$.  
 We combined the ATCA and Parkes data to estimate the flux density of the SNR, because the Parkes data properly includes scales of around 0.5$^{\circ}$ as well, where the ATCA becomes less sensitive.

The vicinity of SNR G353.6-0.7 was observed by the Position Sensitive Proportional Counter (PSPC) on the ROSAT X-ray Observatory during the all-sky survey in 1993 for a total source integration time of 300 seconds (RS932341N00). We obtained an exposure-corrected intensity image of the remnant in the 0.1-2.4 keV band, smoothed with adaptive Gaussian kernel to achieve  Signal-to-noise ratio $\sim$ 5 (see Fig. 1).

\section{Results}
\subsection{Continuum Emission}
The new SNR has a nearly-circular shell-type structure with angular size of 0.5$^{0}$ centred at ({\sl l}, {\sl b})=(353.55, -0.65) in radio continuum (see the first row of Fig. 1). A bright compact HII region G353.42-0.37 is adjacent to the west of the SNR. The 1420 MHz image reveals a clear filament along its south boundary. The 843 MHz image from the Molonglo Observatory Synthesis Telescope (MOST, Green et al. 1999) also shows the filament. The filament of the SNR is also seen at 5 GHz by Parkes observations (Haynes et al. 1978), but not as clearly asat 1420 MHz or 843 MHz due to the lower spatial resolution (4.1$^{prime}$). A few compact sources within the remnant are resolved in both SGPS and MOST maps. Specifically, the bright source G353.45-0.68 is close to the center of the SNR. The SGPS image has a lower resolution (100 arcsec) but higher sensitivity ($<$1~mJy/beam) than the 843~MHz map (43~arcsec and 1 to 2~mJy/beam).  The SGPS image shows more faint emission than the MOST map because the MOST image misses short-spacings more severely than for the SGPS. The 1420~MHz and 843~MHz images have similar shell-type outlines for the SNR. 

21 cm weak polarized emission is seen toward the south filament of SNR G353.6-0.7. 
 However, smoothing the same data with a resolution of 5 arcmin to enhance any extended features shows no convincing polarized emission coming from SNR G353.6-0.7. Absence of observable polarization is consistent with the expectation of complete depolarization at a distance of $> \sim$1 kpc toward the Galactic centre. 

We derive an integrated flux density of 2.2$\pm$0.9 Jy for G353.6-0.7 at 1420 MHz by a new method. We azimuthally averaged the radio intensity in rings around the center of the remnant. Only one half of the
remnant, the part at lowest Galactic latitudes, was used
because the half closest to the Galactic plane is confused
with other sources. The integrated flux density has excluded the contribution from a few compact sources (e.g. G353.45-0.68) within the remnant. Fig. 1 (lower right) shows radio flux as a function of radius of the remnant in diamonds. The background was estimated as the average
intensity at radius $R > 0.4^{\circ}$ and subtracted from the
averaged flux. The profile of the remnant was estimated by a
polynomial. This profile was integrated over radius from
the center of the remnant out to $R = 0.37^{\circ}$ and over
$2\pi$ in angle, which resulted in a total flux density estimate of
$2.2\pm0.9$~Jy. The 843 MHz data misses too many low spatial frequency components to be used to obtain a reliable flux density. However, we obtain an integrated flux density of 1.4$\pm$0.3 Jy at 5 GHz by same above method, giving a spectral index of $\alpha$=0.4$^{-0.6}_{0.5}$ (S$_{\nu} \sim \nu^{-\alpha}$) consistent with a non-thermal feature.   
\begin{figure*}
\vspace{100mm} 
\begin{picture}(80,80)
\put(-20,80){\includegraphics{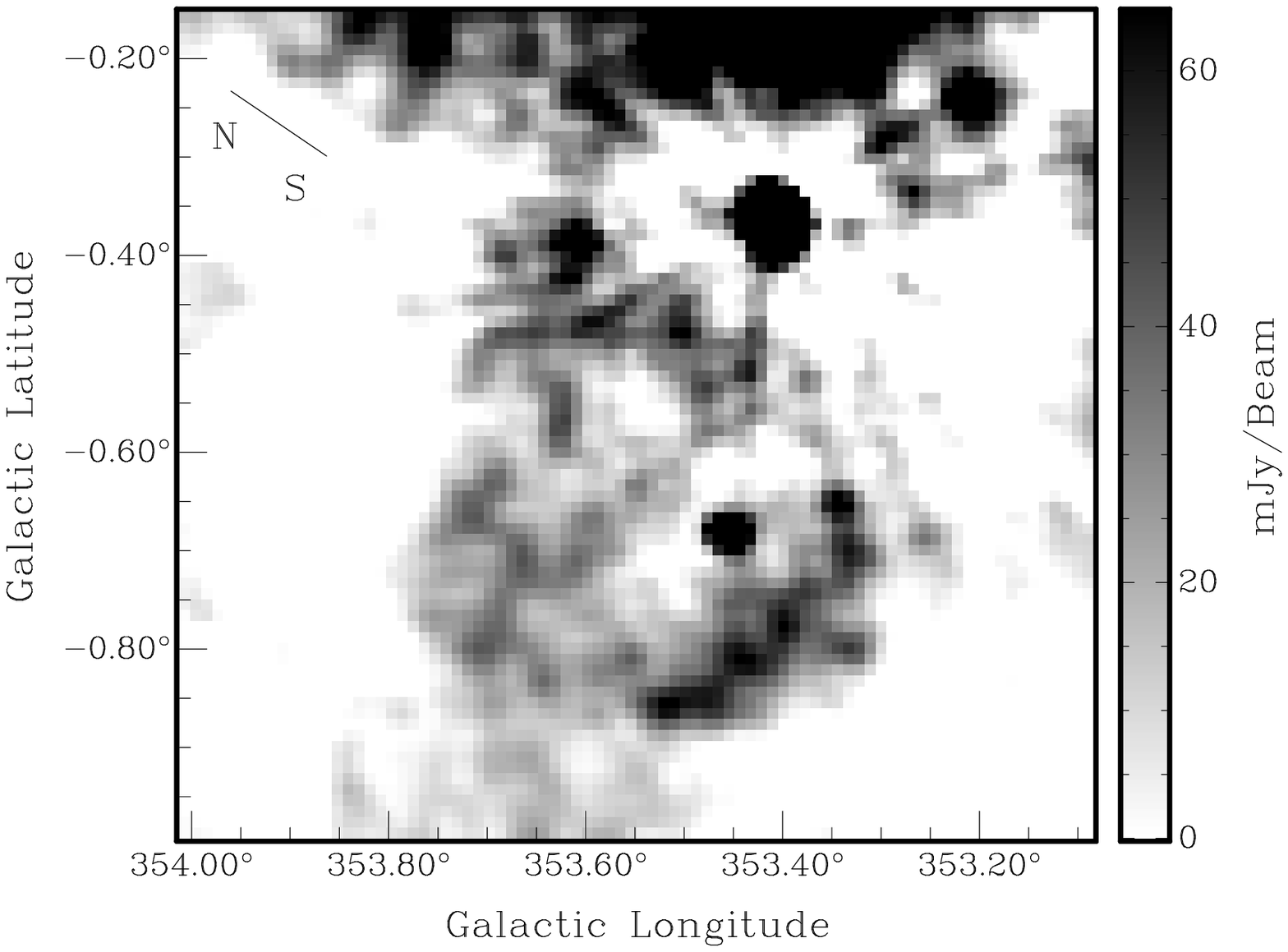}}
\put(215,405){\includegraphics{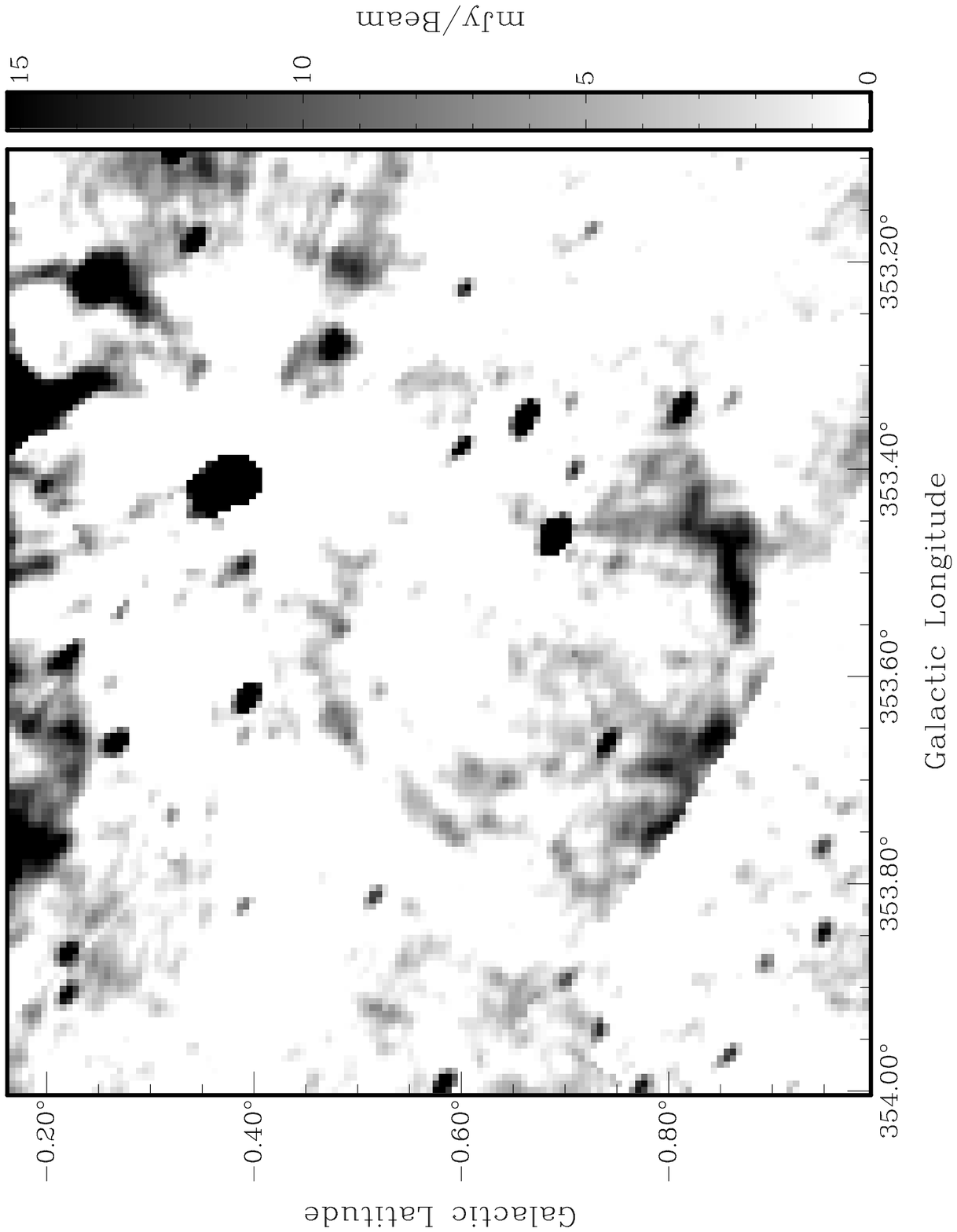}}
\put(-40,220){\includegraphics{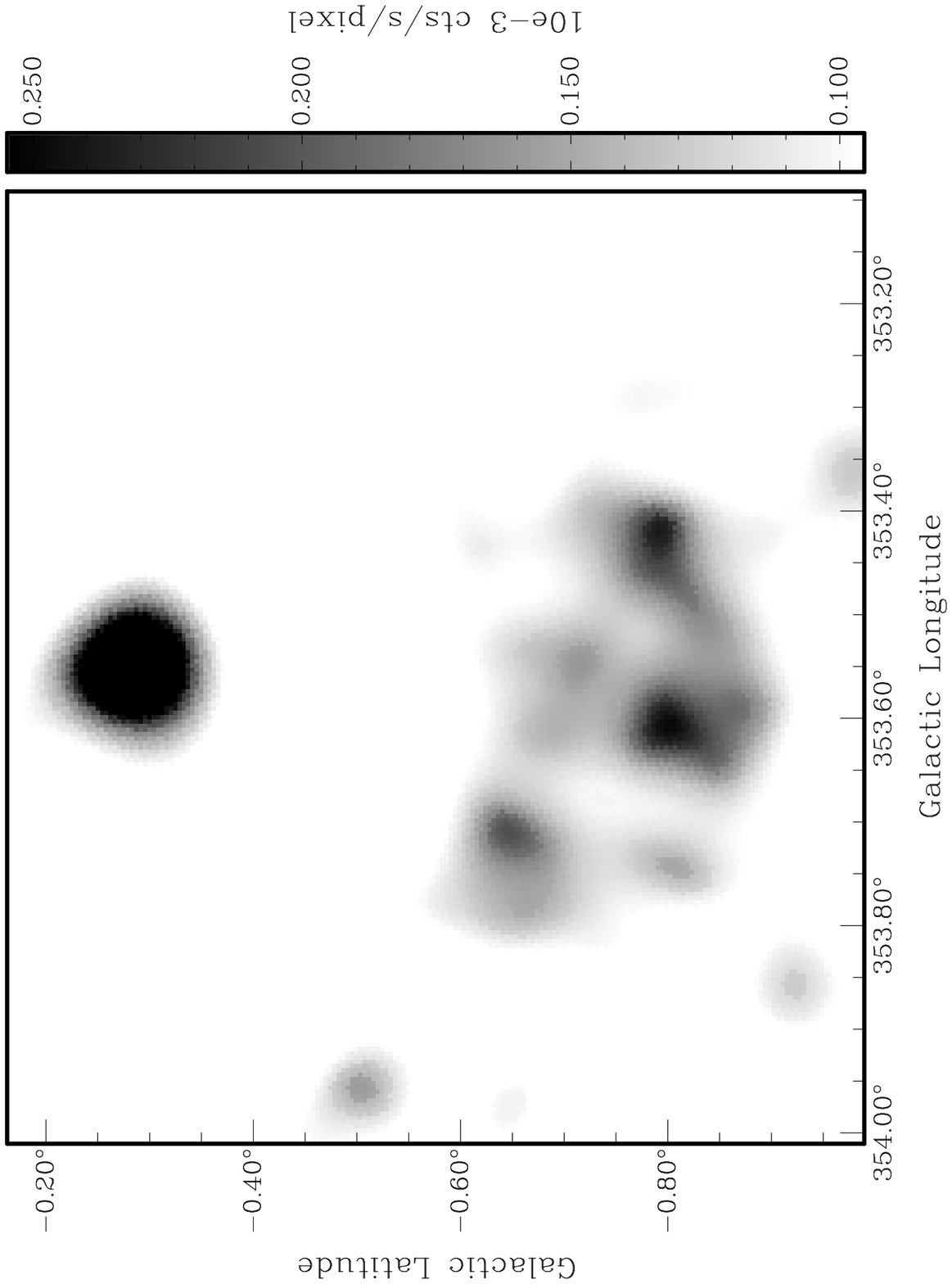}}
\put(235,-160){\includegraphics{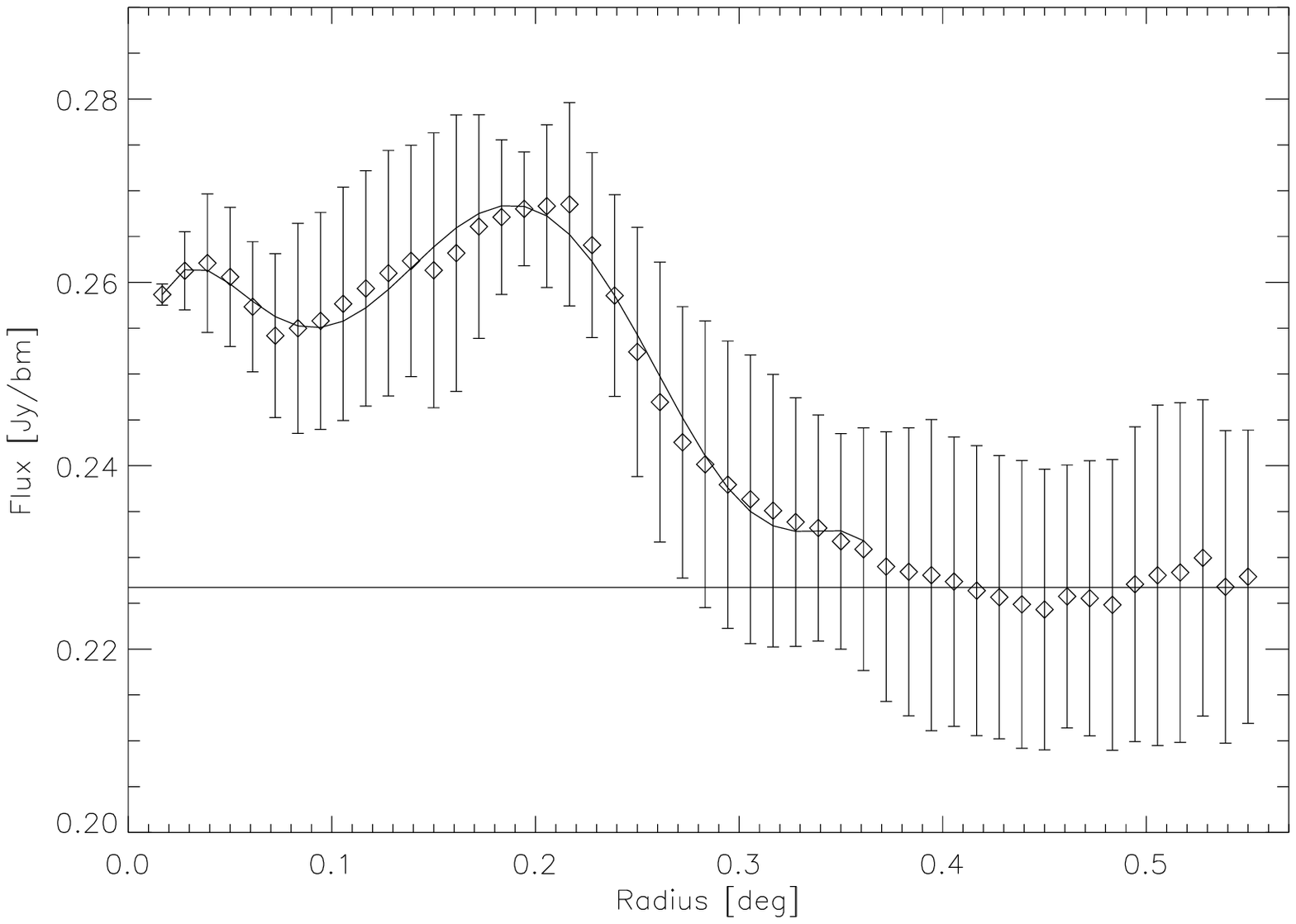}}
\end{picture}
\caption[xx]{The first row shows the continuum images of G353.6-0.7 at 1420 MHz (left) from the SGPS (ATCA data only) with a beamsize of 100~arcsec and at 843 MHz (right) from the Molonglo Galactic Plane Survey with a beamsize of 43~arcsec. The left of the second row presents the ROSAT/PSPC image. The right is azimuthally averaged flux at 1420 MHz over the low-Galactic-latitude half of G353.6-0.7 as a function of radius from the center of the remnant at [b=353.55$^{0}$, l=-0.65$^{0}$]. The horizontal solid line denotes
the background, estimated from the intensity at $R > 0.4^{\circ}$. The
wave solid line is a polynomial fit to the radial profile of the remnant. The direction of North (N) and South (S) is marked on the first plot.} 
\end{figure*} 

\subsection{HI Spectra and Distance Constraints to G353.6-0.7}
The compact HII region G353.42-0.37 is adjacent to the west of the SNR. Fig. 2 shows the HI absorption spectrum features in the range of 9 to -24 km/s which is derived using our improved methods (Tian, Leahy \& Wang 2007; Tian \& Leahy 2008; Leahy \& Tian 2008). Lack of absorption for velocities less than -30 km/s shows G353.42-0.37 is at the near distance (3.2$\pm$0.8 kpc, using a flat Galactic rotation curve model and the most recent estimates of the parameters for this, i.e. velocity V$_{R}$=V$_{0}$=200 km/s and R$_{0}$=7.6 kpc, see Eisenhauer et al. 2005) of its recombination line velocity ($\simeq$ -16 km/s). Based on the fact that a number of HESS sources are detected where high-energy particle accelerators are located near H II regions, and other such accelerators without proximate sources of low-energy photons are weak or undetected at TeV energies, Helfand et al. (2007) suggested that TeV emissions of HESS sources could originate from inverse Compton scattering of the starlight from the nearby H II regions. Given the proximity of bright HII regions to several HESS sources, SNR G353.6-0.7 may be at $\sim$ 3.2 kpc, same as G353.42-0.37. We have also obtained the absorption spectrum of compact source G353.45-0.68, and derived that this source is likely an extragalactic source due to absorption features in the whole negative velocity range. Therefore, we exclude a possibility that G353.45-0.68 is a pulsar or pulsar wind nebula.  


\begin{figure}
\vspace{25mm}
\begin{picture}(80,80)
\put(-20,-10){\includegraphics{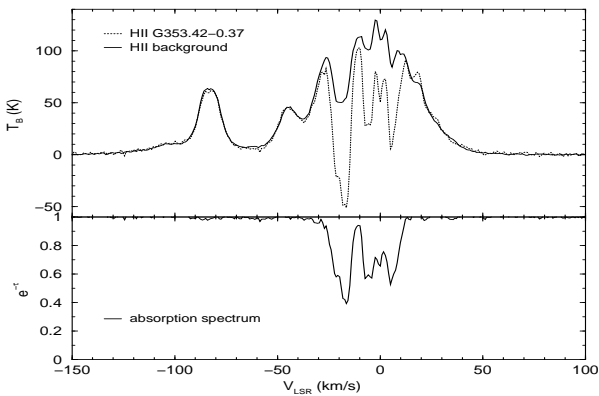}}
\end{picture}
\caption[xx]{The HI spectra of HII G353.42$-$0.37}
\end{figure}

\subsection{Infrared and X-ray Emission}
We analyze new high resolution data (images) at 8$\mu$m from the Galactic Legacy Infrared Midplane Survey Extraordinaire (GLIMPSE) Legacy Project (Benjamin et al. 2003), and a mid-IR image at 21 $\mu$m from the Midcourse Space Experiment (Price et al. 2001). These images (all are centred at [b=353.6, l=-0.7]), show no thermal emission down to a limit of 0.4 mJy (5 $\sigma$) at 8$\mu$m and 200 mJy at 21$\mu$m associated with G353.6-0.7.  This supports above-mentioned conclusion that G353.6-0.7 has a non-thermal emission character and is a new SNR. 

The lower left panel of Fig. 1 shows the ROSAT X-ray image of SNR G353.6-0.7 in the the soft band (0.1-2.4 keV). The diffuse X-ray emission appears only within the lower-latitude half of the radio remnant. The coincidence of the X-ray morphology with the radio morphology (two bright X-ray sources overlap the radio filaments along SNR-shell, and both the X-ray and the radio emission have same shape) suggests that they are physically associated. The fact that X-rays are detected from only the lower half is likely due to absorption blocking X-rays from the upper half. This is
supported by the available low resolution 12CO maps of this area (Dame et al. 2001), showing increasing column density toward the Galactic
plane. 

The estimated unabsorbed X-ray flux of SNR G353.6-0.7 is $\sim$ 8 $\times$ 10$^{-12}$ ergs cm$^{-2}$ s$^{-1}$ in the energy band of 1 - 2 keV (the energy band below 1 keV is affected by absorption) by fitting the ROSAT spectrum with a thermal plasma model (i.e. XSPEC MEKAL with a fitted temperature kT $>$ 2.8 keV). This flux is corresponding to a luminosity of 9.8$\times$ 10$^{33}$ ergs s$^{-1}$ assuming a distance of $\sim$ 3.2 kpc. Due to low number of X-ray counts, this model is quite uncertain. The ROSAT spectrum is extracted from the lower half of SNR G353.6-0.7 (half circle with radius of $15'$) centered on the remnant.

\section{Discussion and Conclusion}
  
\subsection{An old SNR}
 SNR G353.6-0.7 has a large angular size and a large physical size ($\sim$ 28 pc in diameter for a likely distance of $\sim$ 3.2 kpc), and faint radio and X-ray emissions. This suggests that G353.6-0.7 is quite old, because young SNRs ($\le$ 2000 yrs) usually have a size $<$ 10 pc, and produce strong radio and X-ray emission. Applying the Sedov model (Cox 1972), we estimate its age based on the known radius R (14 pc) and ambient density $n_{0}$ assuming a typical explosion energy (10$^{51}$ erg).
The likely absorption of X-rays from the western half of G353.6-0.7 hints that ambient density $n_{0}$ is not very low. The density $n_{0}$ can be estimated from the $\gamma$ flux of HESS J1731-347 ($\it{F_{\gamma}}$$\approx$1.2$\times$10$^{-11}$ cm$^{-2}$ s$^{-1}$ which is given by $dF_{\gamma}/dE(>E_{min}) = N_{0} \left(\frac{E}{1\:{\rm TeV}}\right)^{-\Gamma}$. Here $E_{min}$=0.5 TeV, $\Gamma$=2.26$\pm$0.10, $N_{0}$=6.1$\pm$0.8 $\times$10$^{-12}$ cm$^{-2}$ s$^{-1}$. See Aharonian et al. 2008). Using the formulae 1 and 2 of Aharonian et al. (2006), we obtain the $n_{0}$ of $\sim$ 5 cm$^{-3}$ (assuming a conversion efficiency of 0.1 - 0.2 and explosion energy 10$^{51}$ erg).
If G353.6-0.7 is still in the Sedov-Taylor phase, it has a Sedov age of $\sim$ 25000 yrs.  However, the radius at which cooling affects the dynamics of an SNR is $\sim$ 10 pc for $n_{0}$ $\approx$ 5 cm$^{-3}$, so G353.6-0.7 has already entered the radiative phase. So we here obtain a radiative age of $\sim$ 27000 yrs for G353.6-0.7. 

\begin{figure}
\vspace{45mm}
\begin{picture}(80,80)
\put(322,-15){\includegraphics{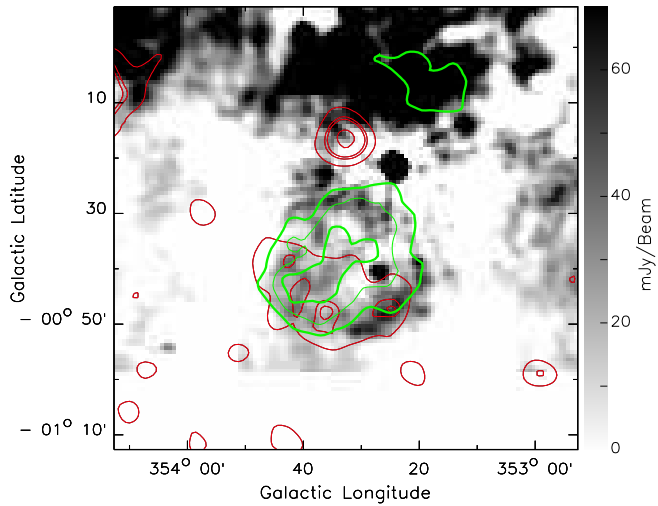}}
\end{picture}
\caption[xx]{The 1420 MHz image of G353.6-0.7 with ROSAT image contours (red: 0.11, 0.19, 0.23, 0.8 in units of 10$^{-3} counts s^{-1} pixel^{-1}$) and HESS J1731-347 image contours (green) overlaid. The HESS J1731-347 image has same contours as Fig. 5 of Aharonian et al. (2008), i.e. starting at $4\sigma $ in $1\sigma $ steps, which is of $\gamma$-ray excess counts smoothed with a Gaussian filter with standard deviation $0.1^\circ $}. 
\end{figure}

\subsection{Radio, X-rays and TeV $\gamma$-rays from G353.6-0.7}
The observations from radio, X-ray and $\gamma$-ray all indicate that SNR G353.6-0.7 is the counterpart of the newly detected TeV $\gamma$-ray source HESS J1731-347 (see Fig. 3). The TeV $\gamma$-rays overlap the entire radio SNR, including the east half seen in X-rays and the west half not seen in X-rays. The $^{12}$CO column density increases toward the Galactic plane, consistent
with absorbing the X-rays from the west half of the SNR, but allowing both radio and TeV $\gamma$-rays to be observed. Old SNRs emitting TeV $\gamma$-rays are rare. W41 has been considered as the only likely candidate so far (Tian et al. 2007). Young SNRs found to show $\gamma$-ray emission have $F_{\gamma(1-10TeV)}/F_{X(2-10 keV)}$ of typically below $\sim$2, $F_{\gamma(1-10TeV)}/F_{radio(0.01-100GHz)}$ of typically below $\sim$10 (Yamazaki et al. 2006).   
  
The flux density of G353.6-0.7 is S$_{1420MHz}$$\approx$ 2.2 Jy, giving an integrated flux from 10$^{7}$ to 10$^{11}$ Hz of F$_{radio}$=5.2 $\times$ 10$^{-13}$ ergs cm$^{-2}$ s$^{-1}$, assuming a radio spectral index of $\sim$ 0.5. We calculate the $\gamma$-ray flux of F$_{\gamma(1-10TeV)}$ $\approx$ 1.7$\times$ 10$^{-11}$ ergs cm$^{-2}$ s$^{-1}$ in the 1-10 TeV band, so the ratio R = F$_{\gamma}$/F$_{radio}$ is $\sim$ 33.
Also given a distance of 3.2 kpc to G353.6-0.7, the $\gamma$-ray luminosity is $\sim$ 2$\times$10$^{34}$ ergs s$^{-1}$ in the 1 - 10 TeV range, which is larger than that of RX J1713.7-3946, one of the intrinsically brightest galactic
 TeV Gamma-ray sources discovered and thought to originate from pion decays. 
So, SNR G353.6-0.7 is a powerful cosmic ray accelerator.

The theoretically predicted $\gamma$-rays to X-rays flux ratio for old SNRs is possibly $> \sim$ 100 (Yamazaki et al. 2006), giving an upper limit to 2-10 keV X-ray emission from G353.6-0.7 of about 1.7$\times$ 10$^{-11}$ ergs cm$^{-2}$ s$^{-1}$.  
We checked the INTEGRAL survey which covers G353.6-0.7 and found no detection within the SNR (JEM-X in energy range of 5 - 20 keV). 
Assuming an individual INTEGRAL observation
for an integration time of 2 ks, the 5 $\sigma$ source detection limit is
1.2$\times$ 10$^{-10}$ ergs cm$^{-2}$ s$^{-1}$ on-axis in the energy range of 2 - 10 keV (Brandt et al. 2003). This is consistent with the derived X-ray upper limit flux from the old SNR.

The fact that the X-rays are visible in SNR G353.6-0.7 shows that SNR shock velocity is still fast. This means that G353.6-0.7 is not too old (i.e. $<$10$^{5}$ yrs), because an old SNR is expected to emit few thermal X-rays which originate from the high temperature gas heated by the fast-enough shock.  Because of poor statistics in the ROSAT X-ray spectrum, we can not exclude that part of the X-rays  are from synchrotron emission. Because no pulsar/PWN is detected within the remnant until now by radio and X-ray observations, it is unlikely that the TeV $\gamma$-rays originate from particles accelerated to high energy at a termination shock formed in the pulsar wind. This strengthens our conclusion that the TeV $\gamma$-rays originate from the old SNR. 

TeV $\gamma$-rays can originate from an old SNR or from a molecular cloud encountered by the SNR, via pion decay with the pions produced by proton-proton collisions (Yamazaki et al. 2006). CO observations can detect all CO cloud components in the direction of G353.6-0.7, therefore help to confirm and distinguish different acceleration mechanisms for emission of TeV $\gamma$-rays. E.g. if there was a CO cloud overlapping the radio, X and $\gamma$-ray emissions at the same distance as SNR G353.6-0.7, and if a broadened CO emission profile was detected, indicating a shock interaction, it would be a strong argument to support the SNR interacting with the cloud to emit TeV $\gamma$-ray emission. However, the existing $^{12}$CO observations in the direction of G353.6-0.7 have too low resolution (1/8 degree) to give any constraint regarding SNR-CO cloud interaction.  
Further high-resolution CO and deep X-ray observations can be used to reveal the real radiative mechanism for TeV emission.

In summary, this new discovery of a radio SNR overlapping the TeV source HESS J1731-347 provides the best case that an old SNR emits TeV $\gamma$-rays.

\begin{acknowledgements}
TWW and DAL acknowledge support from the Natural Sciences and Engineering Research Council of Canada. TWW thanks Dr. Yamazaki for his comments, and the Natural Science Foundation of China for support. The ATCA and Parkes are part of the Australia Telescope, which is funded by the Commonwealth of Australia for operation as a National Facility
managed by CSIRO.  MH acknowledges support from the National Radio Astronomy Observatory, which is operated by Associated Universities Inc.,
under cooperative agreement with the National Science Foundation.   
\end{acknowledgements}


\end{document}